\documentclass[traditabstract,a4]{aa}  

\usepackage{epsfig}
\usepackage{graphicx}
\usepackage{epstopdf}
\usepackage{color}  
\usepackage{array}   
\usepackage{units}  
\usepackage[breaklinks, colorlinks, citecolor=blue]{hyperref}
\usepackage{natbib}  
\usepackage{multirow}   
\usepackage{amsmath,amssymb}  
\usepackage[english]{babel}
\usepackage[latin1]{inputenc}
\usepackage[T1]{fontenc}
\usepackage{longtable,lscape}
\usepackage{footnote}
\usepackage{txfonts}
\usepackage[toc,page]{appendix} 
\begin{document}

%Institutes
\newcommand{\avg}[1]{\left< #1 \right>}

%%%%%%%%%%%%%

%%%%%%%%%%%%%%%
   \title{Constraining galaxy cluster velocity field \\with the tSZ-kSZ-kSZ bispectrum.} 

   \author{G. Hurier\inst{1,2}
          }

\institute{Centro de Estudios de F\'isica del Cosmos de Arag\'on (CEFCA),Plaza de San Juan, 1, planta 2, E-44001, Teruel, Spain
\and Institut d'Astrophysique Spatiale, CNRS (UMR8617) Universit\'{e} Paris-Sud 11, B\^{a}timent 121, Orsay, France
\\
\\
\email{ghurier@cefca.es} 
}

   \date{Received /Accepted}
 
   \abstract{
   The Sunyaev-Zel'dovich (SZ) effects are produced by the interaction of cosmic microwave background (CMB) photons with the ionized and diffuse gas of electrons inside galaxy clusters integrated along the line of sight. The two main effects are the thermal SZ (tSZ) produced by thermal pressure inside galaxy clusters and the kinetic SZ (kSZ) produce by peculiar motion of galaxy clusters compared to CMB rest-frame. The kSZ effect is particularly challenging to measure as it follows the same spectral behavior as the CMB, and consequently can not be separated from the CMB using spectral considerations. In this paper, we explore the feasibility of detecting the kSZ through the computation of the tSZ-CMB-CMB cross-correlation bispectrum for current and future CMB experiments. We conclude that next generation of CMB experiments will offer the possibility to detect at high S/N the tSZ-kSZ-kSZ bispectrum. This measurement will constraints the intra-cluster dynamics and the velocity field of galaxy cluster that is extremely sensitive to the growth rate of structures and thus to dark energy properties. Additionally, we also demonstrate that the tSZ-kSZ-kSZ bispectrum can be used to break the degeneracies between the mass-observable relation and the cosmological parameters to set tight constraints, up to 4\%, on the $Y-M$ relation calibration.}

   \keywords{Cosmology: Observations -- Cosmic background radiation -- Sunyaev-Zel'dovich effect}

   \maketitle

%________________________________________________________________

\section{Introduction}

Present CMB experiments such as ACTT \citep{sie13}, SPT \citep{ble15}, and Planck \citep{planckover} have mapped the cosmic microwave background (CMB) primary temperature anisotropies with an unprecedented precision, and constrained cosmological parameters from the CMB angular power-spectrum analyses \citep[see e.g.,][]{sie13,planckcosmo,haa16}. However, the measurement of secondary anisotropies, such as the Sunyaev-Zel'dovich (SZ) effects \citep{sun72}, is still limited by the noise level and angular resolution in current experiments.

Thanks to it's spectral behavior, the tSZ effect can be isolated from the CMB and foregrounds emissions \citep{rem11,hur13}.
The thermal SZ (tSZ) effect, produced by thermal electron in intra-cluster medium, as been shown as a powerful probe to detect galaxy clusters \citep{has13,ble15,planckpsz2}, and to constrain cosmological parameters using large-scale structure matter distribution \citep{has13,man15,planckszcount,planckszmap,haa16}. However, cosmological constraints obtained via the tSZ effect strongly depend on the mass-observable relation. Assuming a value of $b=0.2$ \citep{planckszcount} for the hydrostatic mass bias, the observed number of galaxy clusters on the sky is only half of the predicted number assuming CMB-derived cosmological parameters \citep{planckcosmo}. CMB-derived cosmological parameter,  favours a hydrostatic mass bias of $b = 0.3-0.5$. Thus, methodology to break the degeneracy between cosmological parameters and the mass-observable relation needs to be investigated further to test the consistency of the $\Lambda$CDM cosmological model.

After the tSZ effect, the second dominant sources of arcminute-scale anisotropies is the kinematic SZ (kSZ) effect \citep{ost86}, produced by the peculiar motion of galaxy clusters with respect to CMB rest-frame. 
The kSZ is about ten times fainter than the tSZ effect and contrary to the tSZ effect, it cannot be separated from the CMB signal using spectral considerations. Consequently, the measurement of this kSZ effect is significantly harder than the one of the tSZ effect. The kSZ effect is directly related to the peculiar velocity field of the matter distribution and the baryonic matter density. Thus, this effect presents significantly different cosmological dependancies than the tSZ effect, especially with the universe growth rate \citep{sug16}, which is a powerful probe to understand the CMB-LSS tension for cosmological parameter estimation.

Several approaches have been proposed to recover the kSZ effect: (i) direct measurement of anisotropies in the CMB angular power spectrum at high-$\ell$ \citep[see][for recent results]{add13, geo15}, (ii) the pairwise momentum estimator, using the preferential motion of large-scale structure toward other large scale structure \citep{pee80, dia00}, (iii) inverting the continuity equation relating density and velocity fields \citep{ho09,kit12}, (iv) cross-correlation bispectrum between CMB and cosmic shear \citep{dor04} or galaxy surveys \citep{ded05}

Evidence of the kSZ effect angular power spectrum has been found using the CMB angular power spectrum \citep{geo15} and combining the tSZ angular power-spectrum and bispectrum \citep{cra14}.
The kSZ detection has been achieve in two cases: for individual galaxy clusters internal dynamics \citep{say13,ada16}, or for statistical sample of galaxy clusters using pairwise momentum \citep{han12,her15,soe16}.
Recent studies \citep[see e.g.,][]{her15} achieve statical detection at about 5 $\sigma$ significance level.

The kSZ effect produces a non-gaussian contribution to the CMB anisotropies. High order statistics have been shown as a powerful probe to detect the tSZ effect \citep[see e.g.,][]{wil12,hil13,planckszmap}.
In this paper, we explore the possibility to detect the tSZ-kSZ-kSZ cross-correlation bispectrum for the next generation of CMB experiments. First, in Sect.~\ref{sec:theory} we present tSZ and kSZ effects, in Sect.~\ref{powspec} we details the computation of tSZ and kSZ power spectra. Then, in Sect.~\ref{bispec} we present the modeling of the tSZ and kSZ bispectra. Finally, in Sect.~\ref{forecast} we present our forecasts for next generation CMB experiments.

\section{The SZ effects}
\label{sec:theory}

The thermal Sunyaev-Zel'dovich effect is a distortion of the CMB black body radiation through inverse Compton scattering. CMB photons receive an average energy boost by collision with hot (a few keV) ionized electrons of the intra-cluster medium \citep[see e.g.][for reviews]{bir99,car02}.
The thermal SZ Compton parameter in a given direction, $\vec{n}$, on the sky is given by
\begin{equation}
y (\vec{n}) = \int n_{e} \frac{k_{\rm{B}} T_{\rm{e}}}{m_{\rm{e}} c^{2} } \sigma_{T} \  \rm{d}s
\label{comppar}
\end{equation}
where $k_{\rm B}$ is the Boltzmann constant, $c$ the speed of light, $m_e$ the electron mass, $\sigma_T$ the Thomson cross-section, $d$s is the distance along the line-of-sight, $\vec{n}$, $n_{\rm{e}}$
and $T_{e}$ are the electron number density and temperature, respectively.
In units of CMB temperature the contribution of the tSZ effect for a given observation frequency $\nu$ is
\begin{equation}
\frac{\Delta T_{\rm{CMB}}}{T_{\rm{CMB}} }= g(\nu) \ y,
\end{equation}
where $T_{CMB}$ is the CMB temperature, and $g(\nu)$ the tSZ spectral dependance.
Neglecting relativistic corrections we have 
\begin{equation}
g(\nu) = \left[ x\coth \left(\frac{x}{2}\right) - 4 \right],
\label{szspec}
\end{equation}
with $ x=h \nu/(k_{\rm{B}} T_{\rm{CMB}})$. At $z=0$, where $T_{\rm CMB}(z=0)$~=~2.726$\pm$0.001~K \citep{mat99}, the tSZ effect is negative below 217~GHz and positive for higher frequencies.\\

The kinetic Sunyaev-Zel'dovich effect produces a shift of the CMB black body radiation temperature but whiteout spectral distortion. Contrary to tSZ effect, it thus posses the same spectral energy distribution (SED) than the CMB primordial anisotropies and cannot be separated from them.
The kinetic SZ induced temperature anisotropies in a given direction on the sky is given by
\begin{equation}
k (\vec{n}) = \int n_{e} \frac{\vec{v} \cdot \vec{n}}{c} \sigma_{T} \  \rm{d}s,
\label{taupar}
\end{equation}
with $\vec{v}$ the peculiar velocity vector of the galaxy cluster. The total kSZ flux from a cluster is proportional to the gas mass, $M_{{\rm gas},500}$ of the cluster, and modulated by the scalar product $\vec{v} \cdot \vec{n}$ that range from $-v$ to $v$.

\section{Power spectra}
\label{powspec}
\subsection{General formalism}
The angular cross power spectrum between two maps reads
\begin{equation}
C^{\rm XY}_\ell = \frac{1}{2\ell +1} \sum_{m} \frac{1}{2}\left(x_{\ell m} y^{*}_{\ell m} + x^{*}_{\ell m} y_{\ell m}\right),
\end{equation}
with $x_{\ell m}$ and $y_{\ell m}$, the coefficients from the spherical harmonics decomposition of the concerned two maps, and $\ell$ the multipole of the spherical harmonic expansion. 
In the context of large scale structure tracers, we model this cross-correlation, as well as the auto correlation power spectra, assuming the following general expression
\begin{equation}
C^{\rm XY}_{\ell} = C^{\rm XY-{\rm 1h}}_\ell + C^{\rm XY-{\rm 2h}}_\ell,
\end{equation}
where $C^{\rm XY-{\rm 1h}}_\ell$ is the Poissonian contribution and $C^{\rm XY-{\rm 2h}}_\ell$ is the 2-halo term. These terms can be computed considering a mass function formalism. The mass function, ${{\rm d^2N}}/{{\rm d}M {\rm d}V}$, gives the number of dark matter halos (in this paper we consider the fitting formula from \citet{tin08}) as a function of the halo mass and redshift.\\

The Poissonian term can be computed by assuming the Fourier transform of normalized halo projected profiles of $X$ and $Y$, weighted by the mass function and the respective fluxes of the halo for $X$ and $Y$ observable \citep[see e.g.][for a derivation of the tSZ auto-correlation angular power spectrum]{col88,kom02}.
{\small
\begin{equation}
C_{\ell}^{\rm XY-{\rm 1h}} = 4 \pi \int {\rm d}z \frac{{\rm d}V}{{\rm d}z {\rm d}\Omega}\int{\rm d}M \frac{{\rm d^2N}}{{\rm d}M {\rm d}V} {X}_{500} {Y}_{500} x_{\ell} y_{\ell},
\end{equation}
}
where ${X}_{500}$ and ${Y}_{500}$ are the average fluxes of the halo in $X$ and $Y$ maps that depend on the critical mass of the galaxy cluster, $M_{500}$, the reshift, $z$, and can be obtained with scaling relations, and ${{\rm d}V}/{{\rm d}z {\rm d}\Omega}$ the comoving volume element. 
%The factor $(1+\sigma_{{\rm log}\, AB})$ accounts for the bias produced by the scatter of scaling relations, and especially correlation between scatter.\\
The Fourier transform of a 3-D profile projected across the line-of-sight on the sphere reads, $\frac{4 \pi r_{\rm s}}{l^2_{\rm s}} \int_0^{\infty} {\rm d}x \, x^2 p(x) \frac{{\rm sin}(\ell x / \ell_{\rm s})}{\ell x / \ell_{\rm s}}$, where $p(x)$ is the halo 3-D profile in $X$ or $Y$ maps, $x = r/r_{\rm s}$, $\ell_{{\rm s}} = D_{\rm ang}(z)/r_{\rm s}$, $r_{\rm s}$ is the scale radius of the profile.\\

The two-halo term corresponds to large scale fluctuations of the dark matter field, that induce correlations in the halo distribution over the sky.
It can be computed as \citep[see e.g.][]{kom99,die04,tab11}
\begin{align}
C_{\ell}^{\rm XY-{\rm 2h}} = 4 \pi \int {\rm d}z \frac{{\rm d}V}{{\rm d}z{\rm d}\Omega}&\left(\int{\rm d}M \frac{{\rm d^2N}}{{\rm d}M {\rm d}V} {X}_{500} x_{\ell} b(M_{500},z)\right)\\ \nonumber
&\times \left(\int{\rm d}M \frac{{\rm d^2N}}{{\rm d}M {\rm d}V} {Y}_{500} y_{\ell} b(M_{500},z)\right) P_{X,Y}(k,z)
\end{align}
with $b(M_{500},z)$ the time dependent linear bias, that relates the power spectrum between $X$ and $Y$ distribution, $P_{X,Y}(k,z)$, to the underlying dark matter power spectrum. 
Following \citet{mo96,kom99} we adopt 
$$b(M_{500},z)=1+(\nu^2(M_{500},z)-1)/\delta_c(z),$$
with $\nu(M_{500},z) = \delta_c(z)/\left[D_g(z) \sigma(M_{500})\right]$, $D_g(z)$ is the linear growth factor and $\delta_c(z)$ is the over-density threshold for spherical collapse.\\

\subsection{The tSZ and kSZ scaling relations}
\label{secscal}
A key step in the modeling of the cross-correlation between tSZ and kSZ is to relate the mass, $M_{500}$, and the redshift, $z$, of a given cluster to tSZ flux, $Y_{500}$, and kSZ flux, $K_{500}$. 
The cross-correlation between tSZ and kSZ effects is thus highly dependent on the $M_{500}-Y_{500}$ and the $M_{500}-K_{500}$ relations in terms of normalization and slope.
Consequently, we need to use the relations derived from a representative sample of galaxy clusters, with careful propagation of statistical and systematic uncertainties. However, such observational constraints are not available for the kSZ effect.
We stress that for power spectrum analysis, the intrinsic scatter of such scaling laws has to be considered, because it will produce extra power that has to be accounted for in order to avoid biases.\\

\begin{table}
\center
\caption{Scaling-law parameters and error budget for both $Y_{500}-M_{500}$ \citep{PlanckSZC} and $Y_{500}-T_{500}$ \citep{PlanckSZC} relations}
\begin{tabular}{|cc|cc|cc|}
\hline
\multicolumn{2}{|c|}{$M_{500}-Y_{500}$} & \multicolumn{2}{|c|}{$M_{500}-T_{500}$} \\
\hline
${\rm log}\,Y_\star$ & -0.19 $\pm$ 0.02 & ${\rm log}\, T_\star$ & -4.27 $\pm$ 0.02   \\
$\alpha_{\rm sz}$ & 1.79 $\pm$ 0.08 & $\alpha_{\rm T}$ & 2.85 $\pm$ 0.18  \\
$\beta_{\rm sz}$ & 0.66 $\pm$ 0.50 &  $\beta_{\rm T}$ & 1 \\
$\sigma_{{\rm log}\, Y}$ & 0.075 $\pm$ 0.010 & $\sigma_{{\rm log}\, T}$ & 0.14 $\pm$ 0.02 \\
\hline
\end{tabular}
\label{tabscal}
\end{table}

We used the $M_{500}-Y_{500}$ scaling laws presented in \citet{PlanckSZC}, 
\begin{equation}
E^{-\beta_{\rm sz}}(z) \left[ \frac{D^2_{\rm A}(z) {Y}_{500}}{10^{-4}\,{\rm Mpc}^2} \right] = Y_\star \left[ \frac{h}{0.7} \right]^{-2+\alpha_{\rm sz}} \left[ \frac{(1-b) M_{500}}{6 \times 10^{14}\,{\rm M_{\odot}}}\right]^{\alpha_{\rm sz}},
\label{szlaw}
\end{equation}
with $h$ the dimensionless Hubble parameter, $D_{\rm A}$ the angular distance, $E(z) = \Omega_{\rm m}(1+z)^3 + \Omega_{\Lambda}$. The coefficients $Y_\star$, $\alpha_{\rm sz}$, and $\beta_{\rm sz}$ from \citet{PlanckSZC}, are given in Table~\ref{tabscal}. We used $b=0.2$ for the bias between X-ray estimated mass and effective mass of the clusters.

We also need to have an estimate of the cluster temperature, ${T}_{500}$. In this work, we used the scaling law from \citet{planckSL}:
\begin{equation}
E(z)^{-\beta_{\rm T}} {Y}_{500} = T_{\star} \left[ \frac{{T}_{500}}{6\,{\rm keV}} \right]^{\alpha_{\rm T}},
\end{equation}
where the coefficients $T_\star$, $\alpha_{\rm T}$ and $\beta_{\rm T}$ are given in Table~\ref{tabscal}.

To model the $K_{500}-M_{500}$ relation, we consider the relation from \citet{ded05} for the velocity field:
\begin{align}
v_k = i\frac{H(z) D_g}{1+z} \frac{{\rm d \, ln} (D_g)}{{\rm d \, ln}(a)} \frac{\delta_k}{k},
\end{align}
and consequently for the velocity dispersion
\begin{align}
\sigma_v(M_{500},z) =  \frac{H(z) D_g}{1+z} \frac{{\rm d \, ln} (D_g)}{{\rm d \, ln}(a)} \sigma_{-1}, %\sqrt{1-\frac{\sigma^4_{0}}{\sigma^2_{-1}\sigma^2_{1}}},
\end{align}
with $D_g$ the growth factor and $\sigma_{j}(z)$ is defined for any integer $j$ as
\begin{align}
\sigma_j(M_{500}) = \frac{1}{2 \pi^2} \int {\rm d}k \, k^{2(1+j)} P(k) W^2(kR),
\end{align}
where $W(kR) = \frac{3}{(kR)^3}\left( {\rm sin}(kR) - kR \, {\rm cos}(kR) \right)$ is the Fourier transform of he real space top-hat window function, with $R = \sqrt{\frac{3M_{500}}{4 \pi \bar{\rho}}}$, where $\bar{\rho}$ is the critical density of the universe.

%\begin{equation}
%\left[ \frac{D^2_{ang}(z) {K}_{500}}{10^{-4}\,{\rm Mpc}^2} \right] = K_\star \sigma_v(M,z) \left[ \frac{ (1-b) M_{500}}{6 \times 10^{14}\,{\rm M_{\odot}}}\right],
%\end{equation}
By simplicity, to estimate the kSZ flux root-mean-square (over velocities), $K_{500}$, we consider the relation
\begin{align}
{Y}_{500} \simeq {K}_{500} \frac{ k_{\rm B}{T}_{500}}{m_e c^2} \left( \frac{\sigma_v}{c}\right)^{-1}.
\end{align}
$K_{500}$ is proportional to $M_{{\rm gas},500}$, thus the $K_{500}-M_{500}$ and $M_{{\rm gas},500}-M_{500}$ intrinsic scatters are the identical. The extra-scatter induced by the velocities is accounted by the $\sigma_v$ factor. We derived a $K_{500}-M_{500}$ intrinsic scatter of $\sigma_{{\rm log}\, K} = 0.03 \pm 0.01$ from the galaxy cluster sample presented in \citet{planckSL}.\\
Additionally, the $Y_{500}-M_{500}$ and $K_{500}-M_{500}$ intrinsic scatters are correlated. The intrinsic scatters follow the relation
\begin{align}
\sigma^2_{{\rm log}\, YK} = \sigma^2_{{\rm log}\, Y} + \sigma^2_{{\rm log}\, K} - 2 \rho \sigma_{{\rm log}\, Y} \sigma_{{\rm log}\, K},
\end{align}
where $\sigma_{{\rm log}\, YK} = 0.08 \pm 0.01$ is the intrinsic scatter of the $Y_{500}-K_{500}$ (equivalent to the scatter of the $Y_{500}-M_{{\rm gas},500}$) and $\rho$ is the correlation factor between $Y_{500}-M_{500}$ and $K_{500}-M_{500}$ intrinsic scatters. We derived $\rho \in [0.5,1.0]$\footnote{For consistency, we used $\sigma_{{\rm log}\, Y} = 0.10 \pm 0.01$ that has been derived on the galaxy cluster sample from \citet{planckSL}}. In the following we assume $\rho = 0.75$.

\subsection{Log-normal scatter and n-points correlation functions}

Scaling relations shown in sect.\ref{secscal} presents an intrinsic physical scatter (see tab.~\ref{tabscal}).
This scatter is generally considered as a log-normal distribution. When entering in correlation functions, this log-normal scatter will act as an additional source of power. This additional power, for a quantity $X$, can be expressed as a relation between $<X^n>$ and $X_\star$, where $X_\star$ is the log-normal mean of the $X$ variable. In the most general case, the $n$-th momentum, $M^{(n)}$, expectation of a set of $N$ variable $X_i$ can be written
\begin{align}
M^{(n)} &= <\prod^N_i X^{n_i}_i> \nonumber \\
<M^{(n)}> &= A_{\rm norm} \int \prod^N_i X^{n_i}_i {\rm d}X_i \, \nonumber\\
&{\rm exp}\left( - \frac{\left[ {\rm log}({\bf X}) - {\rm log}({\bf X_\star})\right]^{\rm T} {\cal C}^{-1}_{\rm S} \left[  {\rm log}({\bf X}) - {\rm log}({\bf X_\star})\right]}{2} \right),
\end{align}
where $ A_{\rm norm}$ is the normalisation factor of the log-normal distribution, ${\cal C}_{\rm S}$ is the scaling relation scatter covariance matrix, ${\bf X}$ is a vector of $X_i$ variables, ${\bf X_\star}$ is a vector containing the log-normal expectation $X_{\star,i}$ for each variable $X_i$, and $n_i$ is the order of each variable $X_i$ in the momentum $M^{(n)}$ and satisfy ${\sum^N_i n_i = n}$.
It can be easily shown that
\begin{align}
<M^{(n)}> = {\rm exp} \left( \frac{{\bf n}^{\rm T} {\cal C}_{\rm S} {\bf n} }{2}\right) \, \prod_i^N X^{n_i}_{\star,i},
\end{align}
where ${\bf n}$ is a vector of $n_i$.
This effect produce a power enhancement of 6\% for the tSZ bispectrum. Which is significant for high-S/N measurement of tSZ bispectrum. For the the tSZ-kSZ-kSZ bispectrum, we derived a power enhancement of 1.8\%.
In the following, we correct all spectra for this effect.  

\subsection{Pressure and density profiles}
\label{secprof}

The tSZ effect is directly proportional to the pressure integrated across the line of sight. In this work, we model the galaxy cluster pressure profile by a generalized Navarro Frenk and White \citep[GNFW,][]{nav97,nag07} profile of the form
\begin{equation}
{\mathbb P}(r) = \frac{P_0}{\left(c_{500} r\right)^\gamma \left[1 + (c_{500} r)^\alpha \right]^{(\beta-\gamma)/\alpha}}.
\end{equation}
For the parameters $c_{500}$, $\alpha$, $\beta$, and $\gamma$, we used the best-fitting values from \citet{arn10} presented in Table.~\ref{tabscal}. The absolute normalization of the profile $P_0$ is set assuming the scaling laws $Y_{500}-M_{500}$ presented in Sect.~\ref{secscal}.\\
To model the kSZ profile, we need the density, $n_e(r)$, profile. Thus, we assume a polytropic equation of state \citep[see, e.g.,][]{kom01}, ${\mathbb P}(r) = n_e(r)T_e(r)$, with
$n_e(r) \propto T_e(r)^\delta$ where $\delta$ is the polytropic index.
Considering that the kSZ varies with $n_e$, the kSZ profile is proportional to $P(r)^{\epsilon_{p}}$, where $\epsilon_p = \frac{\delta}{\delta+1}$ ranges from 0.5 to 1.0 for $1.0 <\delta< \infty$. \\ 
The overall normalization of kSZ profile is deduced from the scaling law $K_{500}-M_{500}$ presented in Sect.~\ref{secscal}.

\subsection{tSZ and kSZ power spectra}

\renewcommand{\arraystretch}{1.5}
\begin{table}
\center
\caption{Amplitude of the different terms contribution to the tSZ and kSZ power spectra and bispectra. $P_{\rm B}$ is the power spectrum of the momentum field and $P_{\rm m}$ the matter power spectrum.}
\begin{tabular}{|c|c|c|c|}
\hline
& 1-halo & 2-halo & 3-halo \\
\hline
$C_{\ell}^{\rm tSZ-kSZ}$ & 0 & 0 & \\
$C_{\ell}^{\rm kSZ-kSZ}$ & 1/3 & $P_{\rm B}$ & \\
\hline
$b_{\ell_1 \ell_2 \ell_3}^{\rm tSZ-tSZ-kSZ}$ & 0 & (0,0,0) & 0 \\ 
$b_{\ell_1 \ell_2 \ell_3}^{\rm tSZ-kSZ-kSZ}$ & 1/3 & ($P_{\rm B}$,$P_{\rm B}$,$P_{\rm m}$/3) & $B_k$ \\
$b_{\ell_1 \ell_2 \ell_3}^{\rm kSZ-kSZ-kSZ}$ & 0 & (0,0,0) & 0 \\
\hline
\end{tabular}
\label{tabfact}
\end{table}
\renewcommand{\arraystretch}{1}

\begin{figure}[!th]
\begin{center}
\includegraphics[scale=0.2,trim = 0cm 0cm 0cm 1.3cm, clip]{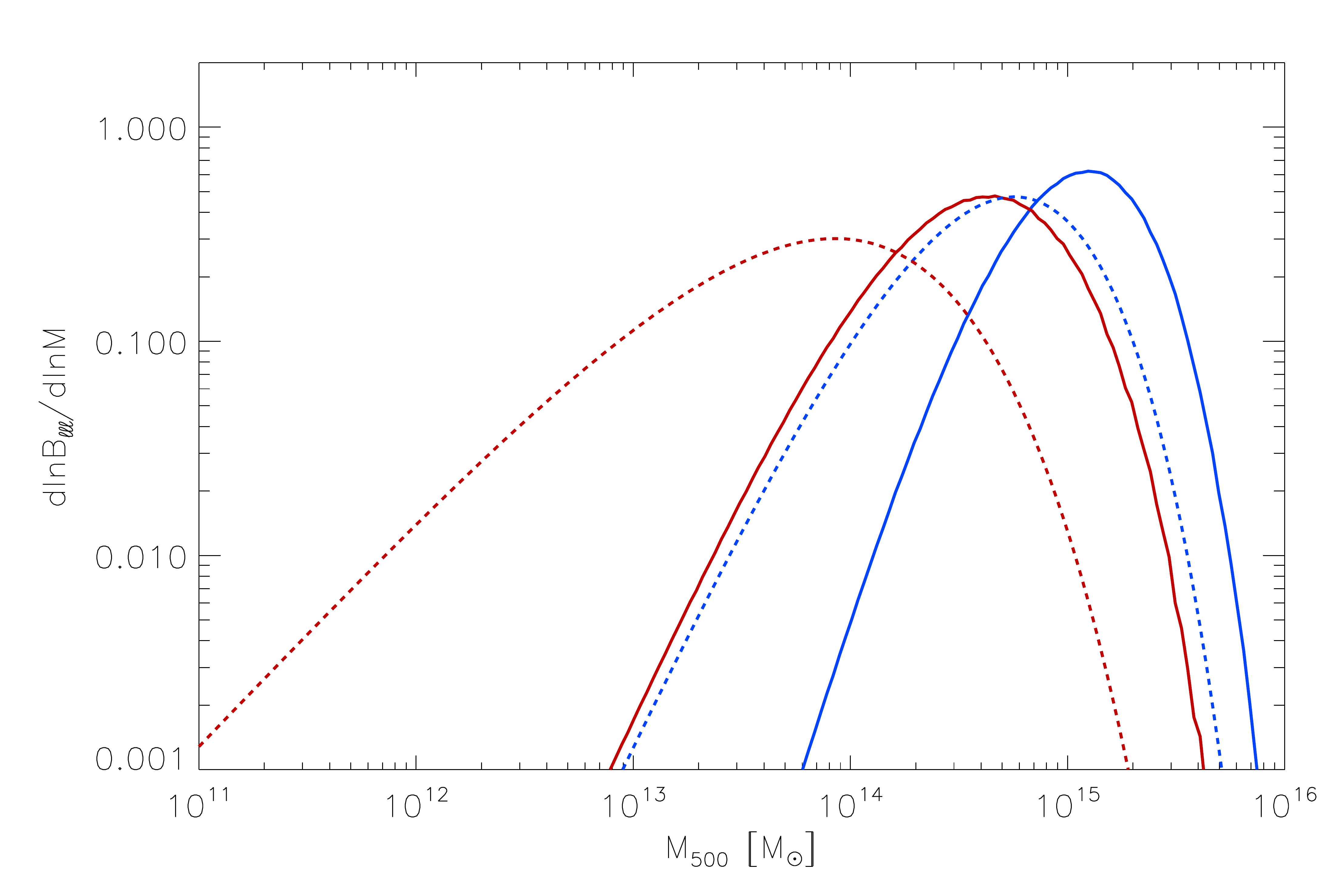}
\includegraphics[scale=0.2,trim = 0cm 0cm 0cm 1.3cm, clip]{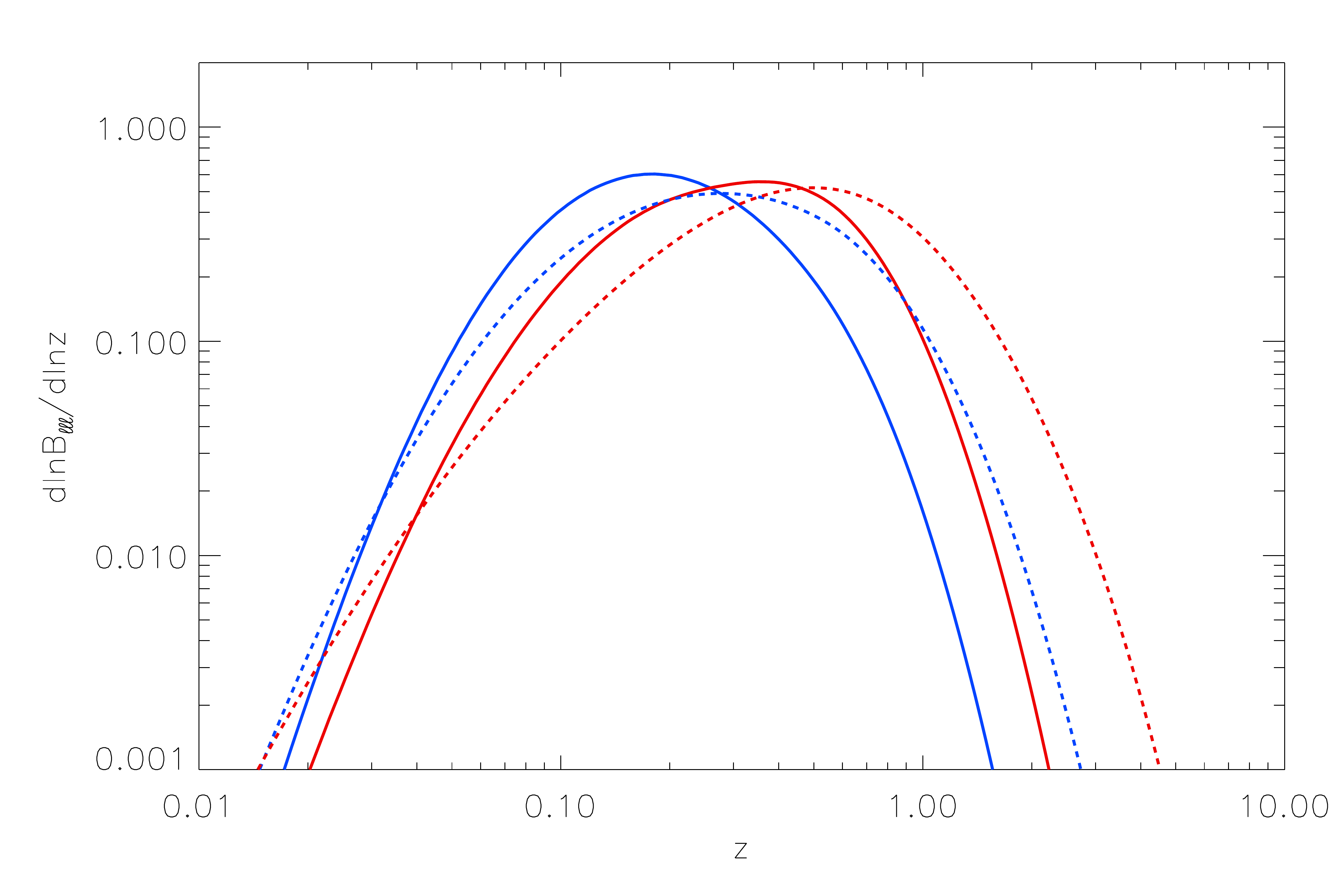}
\caption{Power density as a function of $M_{500}$ (top panel) and redshift (bottom panel) for tSZ power spectrum (dotted blue line), tSZ bispectrum (solid blue line), kSZ power spectrum (dotted red line), and tSZ-kSZ-kSZ bispectrum (solid red line).}.
\label{figdistrib}
\end{center}
\end{figure}

The kSZ effect is dependent of the orientation of the peculiar velocity vector. Consequently, power spectra have to be averaged over orientations.
In Table~\ref{tabfact}, we present the multiplicative factors to be applied on kSZ related power spectra, where $P_{\rm B}$ is the power spectrum of the momentum field and $P_{\rm m}$ is the power spectrum of the matter field. We notice that kSZ effect present a 2-halo term, induce by the large scale correlations between the velocities of different clusters.

On Fig.~\ref{figdistrib}, we present the power density as a function of $M_{500}$ and $z$ for tSZ and kSZ power spectra, tSZ bispectrum, and tSZ-kSZ-kSZ cross-bispectrum.
We observe that tSZ power spectrum samples higher mass and lower redshift than kSZ power spectrum, this effect is a consequence of the slope of $Y_{500}-M_{500}$ (~1.7) and $K_{500}-M_{500}$ (~1). We also observe that bispectra gives more weights to very massive and nearby objects.

\section{bispectrum}
\label{bispec}
Combining tSZ and kSZ, it is possible to build two auto-correlation bispectra, and two cross-correlation bispectra.
In this work we aim at predicting the tSZ-kSZ-kSZ bispectrum, thus our modeling only accounts for non-gaussian objects the correlate with the tSZ effect. Consequently, we only consider the kSZ effect from virialized halos and we neglect the kSZ contribution produced by diffuse baryons at higher redshift.

\subsection{cross-correlation bispectra}

Following the same halo model approach, we can easily predict the SZ bispectra \citep[see][for a detailed description of the tSZ bispectrum]{bha12}.
In halo model formalism, a bispectrum can be separated into 3 terms: one-halo, two-halo, and three-halo, as
\begin{equation}
b_{\ell_1 \ell_2 \ell_3}^{\rm XYZ} = b_{\ell_1 \ell_2 \ell_3}^{\rm XYZ-{\rm 1h}} + b_{\ell_1 \ell_2 \ell_3}^{\rm XYZ-{\rm 2h}} + b_{\ell_1 \ell_2 \ell_3}^{\rm XYZ-{\rm 3h}},
\end{equation}

The one-halo term, is produce by the auto-correlation of a cluster with himself,
{\small
\begin{equation}
b_{\ell_1 \ell_2 \ell_3}^{\rm XYZ-{\rm 1h}} = 4 \pi \int {\rm d}z \frac{{\rm d}V}{{\rm d}z {\rm d}\Omega}\int{\rm d}M \frac{{\rm d^2N}}{{\rm d}M {\rm d}V} {X}_{500} {Y}_{500} {Z}_{500} x_{\ell_1}y_{\ell_2}z_{\ell_3},
\end{equation}
}

The two-halo involved two point from the same halo and the third from another one. As a consequence, this term receives three contributions,
{\small
\begin{align}
b_{\ell_1 \ell_2 \ell_3}^{\rm XYZ-{\rm 2h}} = & 4 \pi \int {\rm d}z \frac{{\rm d}V}{{\rm d}z{\rm d}\Omega}P_{XY,Z}(k,z)\left(\int{\rm d}M \frac{{\rm d^2N}}{{\rm d}M {\rm d}V} {X}_{500} {Y}_{500} x_{\ell_1}y_{\ell_2} b(M_{500},z)\right) \nonumber \\ \nonumber
&\times \left(\int{\rm d}M \frac{{\rm d^2N}}{{\rm d}M {\rm d}V} {Z}_{500} z_{\ell_3} b(M_{500},z)\right) \\ \nonumber
&+ 4 \pi \int {\rm d}z \frac{{\rm d}V}{{\rm d}z{\rm d}\Omega}P_{XZ,Y}(k,z)\left(\int{\rm d}M \frac{{\rm d^2N}}{{\rm d}M {\rm d}V} {X}_{500}{Z}_{500} x_{\ell_1}z_{\ell_3} b(M_{500},z)\right)\\ \nonumber
&\times \left(\int{\rm d}M \frac{{\rm d^2N}}{{\rm d}M {\rm d}V} {Y}_{500} y_{\ell_2} b(M_{500},z)\right) \\ \nonumber
&+ 4 \pi \int {\rm d}z \frac{{\rm d}V}{{\rm d}z{\rm d}\Omega}P_{YZ,X}(k,z)\left(\int{\rm d}M \frac{{\rm d^2N}}{{\rm d}M {\rm d}V} {Y}_{500}{Z}_{500} y_{\ell_2}z_{\ell_3} b(M_{500},z)\right)\\ \nonumber
&\times \left(\int{\rm d}M \frac{{\rm d^2N}}{{\rm d}M {\rm d}V} {X}_{500} x_{\ell_1} b(M_{500},z)\right) \\ 
\end{align}
}

The tree-halo involved the correlation of three different halos,
{\small
\begin{align}
b_{\ell_1 \ell_2 \ell_3}^{\rm XYZ-{\rm 3h}} =  4 \pi \int {\rm d}z \frac{{\rm d}V}{{\rm d}z{\rm d}\Omega}B_{X,Y,Z}(k_1,k_2,k_3,&z)\left(\int{\rm d}M \frac{{\rm d^2N}}{{\rm d}M {\rm d}V} {X}_{500}  x_{\ell_1} b_3(M_{500},z)\right) \nonumber \\
&\times \left(\int{\rm d}M \frac{{\rm d^2N}}{{\rm d}M {\rm d}V} {Y}_{500} y_{\ell_2} b_3(M_{500},z)\right) \nonumber \\
& \times \left(\int{\rm d}M \frac{{\rm d^2N}}{{\rm d}M {\rm d}V} {Z}_{500} z_{\ell_3} b_3(M_{500},z)\right) 
\end{align}
}
with $B_{X,Y,Z}(k_1,k_2,k_3,z)$ the bispectrum of $X$, $Y$, and $Z$ distribution and $b_3(M_{500},z)$ the bias that relates dark-matter and halo distributions.

\subsection{The tSZ-kSZ-kSZ cross-bispectrum}

\begin{figure}[!th]
\begin{center}
\includegraphics[scale=0.2,trim = 0cm 0cm 0cm 1.3cm, clip]{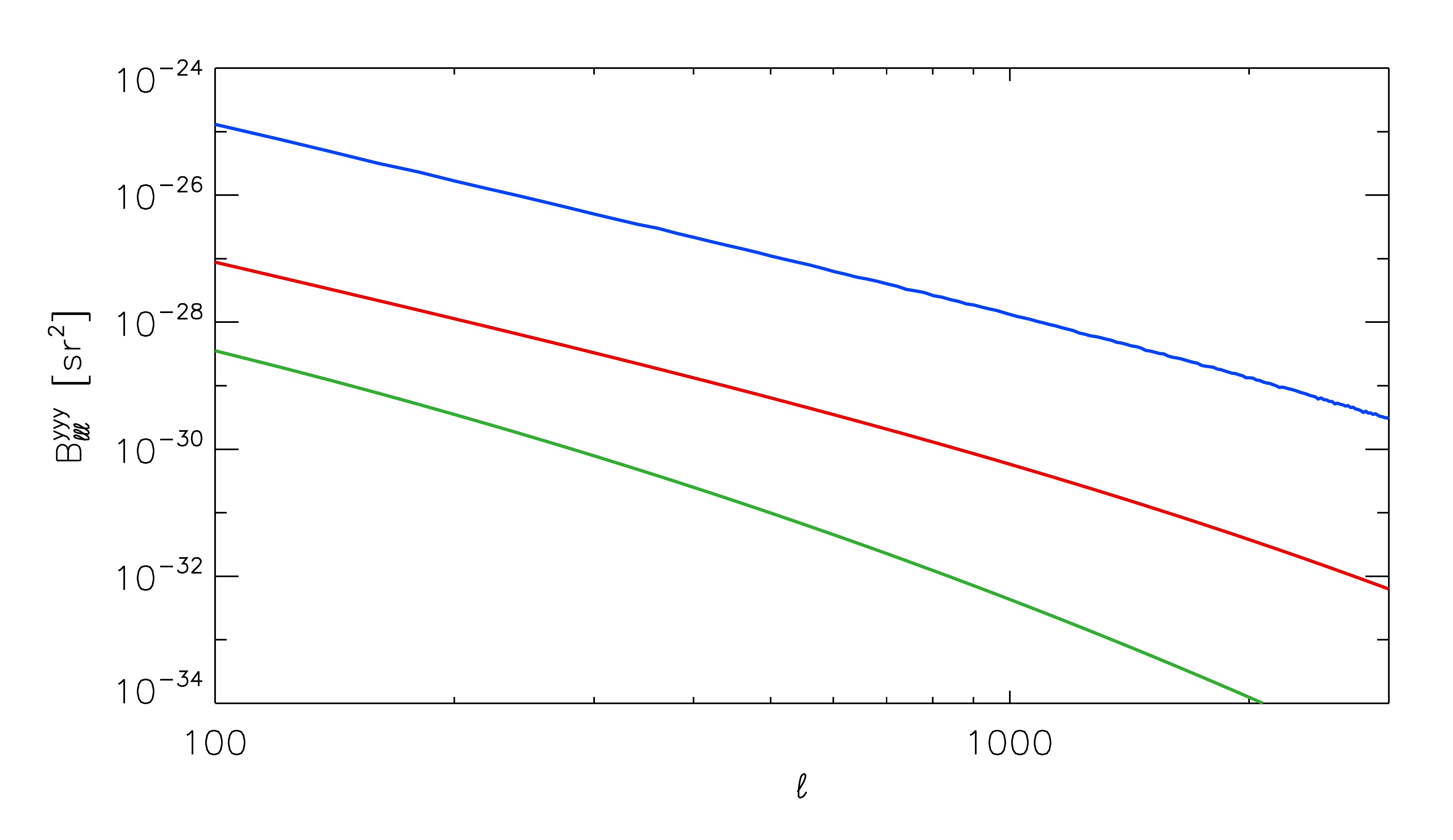}
\includegraphics[scale=0.2,trim = 0cm 0cm 0cm 1.3cm, clip]{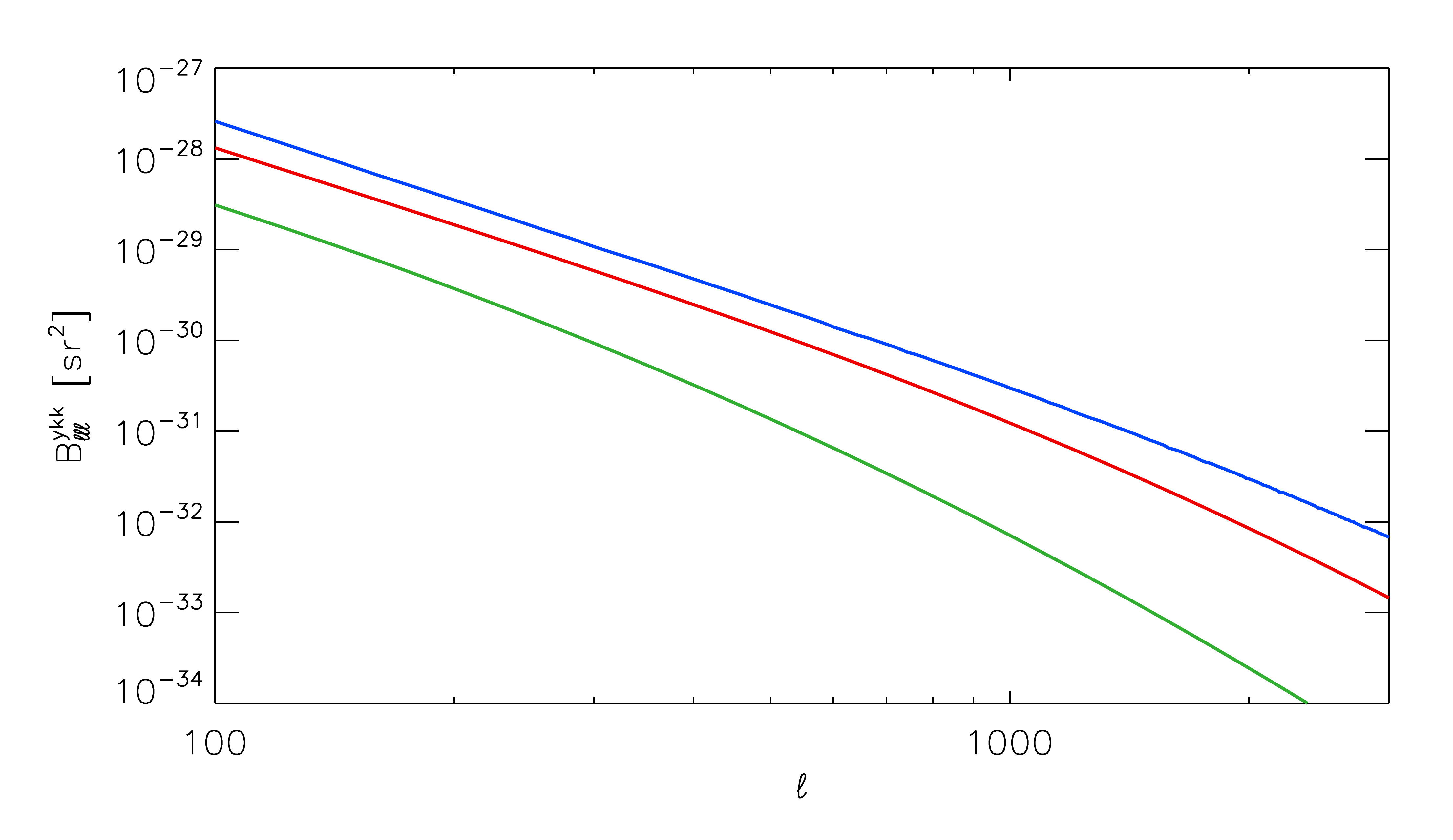}
\caption{tSZ (top panel) and tSZ-kSZ-kSZ (bottom panel) bispectra 1 halo term (blue line), 2-halo term (red line), and 3-halo term (green line).}.
\label{figbl}
\end{center}
\end{figure}

The tSZ-tSZ-kSZ bispectrum have a null expectation due to the average over all the direction for the momentum field of galaxy clusters. Table~\ref{tabfact} lists the relative amplitude of each contributions.\\
For the prediction of tSZ-kSZ-kSZ 2-halo term, we compute the momentum field power spectrum similarly to \citet{sha12}. 
The tSZ-kSZ-kSZ bispectrum 2-halo term involves three contributions. Two contributions where the two considered halos receive contribution from the kSZ effect. These terms involve the momentum field power spectrum. The third contributions involves the correlation between a halo weighted by his tSZ flux and a halo weighed by the square of the kSZ effect. Consequently, this third contribution involves the matter power spectrum that describes the distribution of halos over the sky, weighted by the average over momentum directions (similarly to the kSZ power spectrum).
For the 3-halo term matter-momentum bispectrum we used the prescription from \citep{ded05}.\\

We studied the cosmological parameter dependancies of this cross bispectrum. We found that the tSZ-kSZ-kSZ bispectrum is proportional to $\Omega^{4.1}_m \sigma^{10.1}_8 H^{2.0}_0$, by comparison the tSZ bispectrum is proportional to $\Omega^{3.9}_m \sigma^{12.9}_8 H^{-1.1}_0$ et $\ell \simeq 1000$. Then, tSZ-kSZ-kSZ might be a powerful probe to break degeneracies between cosmological parameters and scaling relations by providing similar cosmological dependancies than tSZ power spectrum with significantly different astrophysical processes dependancies. We also note that the tSZ bispectrum present similar degeneracies than the tSZ power spectrum, consequently it cannot be used to break degeneracies between cosmology and astrophysical processes.\\

On Fig.~\ref{figdistrib}, we present the power density for tSZ and tSZ-kSZ-kSZ bispectra. We observe that bispectra are sensitive to higher mass and lower redshift than power spectra. We note that tSZ-kSZ-kSZ bispectrum and tSZ power spectrum presents similar power density distribution as a function of mass and redshift. the tSZ-kSZ-kSZ bispectrum is thus sensitive to galaxy clusters with mass ranging from $10^{14}$ and $10^{15}\, M_\odot$, at $z < 1$.

We present the tSZ and tSZ-kSZ-kSZ bispectra on Fig.~\ref{figbl}. We observed that for the tSZ bispectrum 1 halo term is two order of magnitudes higher then 2 two halo. Consequently, for the tSZ bispectrum 2-halo and 3-halo terms can be safely neglected. This is consistent with the tSZ angular power spectrum that only presents significant contribution from the 2-halo term at very low-$\ell$. When using higher order statistics we favours higher mass, lower-$z$ objects that are less frequent over the sky. Indeed, the 1-halo term amplitude evolved with the number, $N_{\rm cl}$, that significantly contributes to the spectra, where the 2-halo term evolves as $N^2_{\rm cl}$ and the 3-halo term as $N^3_{\rm cl}$.\\
Contrary to the tSZ bispectrum, the tSZ-kSZ-kSZ bispectrum 2-halo and 3-halo term present significant contributions compared to the 1-halo terms. This higher contribution is explained by the fact that the kSZ effect favours lower-mass and higher-z objects than the tSZ effect, as shown on Fig.~\ref{figdistrib}.

\subsection{uncertainties and optimal estimator}

To compute the bispectrum between three maps $X$, $Y$, and $Z$, we considered the following formula
\begin{align}
b_{\ell_1\ell_2\ell_3} = \frac{1}{f_{\rm sky}}\sum_{\vec{n}} & X_{\ell_1} Y_{\ell_2} Z_{\ell_3} ,
\end{align}
where $X_{\ell_1}$, $Y_{\ell_2}$, $Z_{\ell_3}$ are the real space map that only contain $\ell_1$, $\ell_2$, $\ell_3$ multipoles of maps $X$, $Y$, $Z$ respectively, $f_{\rm sky}$ is the covered sky fraction, and $\vec{n}$ is the direction over the sky.\\
This estimator is known to reduce the variance of the bispectrum without biasing the expectation of the bispectrum.
We note that, in the case of a cross-correlation bispectrum $\ell_1$, $\ell_2$, $\ell_3$ are not commutative quantities.

Then the bispectrum variance in the weak non-Gaussianity limit can be expressed as
\begin{align}
\label{eqvar}
<b_{\ell_1\ell_2\ell_3},b_{\ell'_1\ell'_2\ell'_3}> =& \frac{C^{\rm XX}_{\ell_1}C^{\rm YY}_{\ell_2}C^{\rm ZZ}_{\ell_3}}{f_{\rm sky}N_{\ell_1,\ell_2,\ell_3}}\delta_{\ell_1 \ell'_1}\delta_{\ell_2 \ell'_2}\delta_{\ell_3 \ell'_3} \nonumber \\
&+\frac{C^{\rm XX}_{\ell_1}C^{\rm YZ}_{\ell_2}C^{\rm ZY}_{\ell_3}}{f_{\rm sky}N_{\ell_1 \ell_2 \ell_3}}\delta_{\ell_1 \ell'_1}\delta_{\ell_2 \ell'_3}\delta_{\ell_3 \ell'_2} \nonumber \\
&+\frac{C^{\rm XY}_{\ell_1}C^{\rm YX}_{\ell_2}C^{\rm ZZ}_{\ell_3}}{f_{\rm sky}N_{\ell_1 \ell_2 \ell_3}}\delta_{\ell_1 \ell'_2}\delta_{\ell_2 \ell'_1}\delta_{\ell_3 \ell'_3} \nonumber \\
&+\frac{C^{\rm XY}_{\ell_1}C^{\rm YZ}_{\ell_2}C^{\rm ZX}_{\ell_3}}{f_{\rm sky}N_{\ell_1 \ell_2 \ell_3}}\delta_{\ell_1 \ell'_2}\delta_{\ell_2 \ell'_3}\delta_{\ell_3 \ell'_1} \nonumber \\
&+\frac{C^{\rm XZ}_{\ell_1}C^{\rm YY}_{\ell_2}C^{\rm ZX}_{\ell_3}}{f_{\rm sky}N_{\ell_1 \ell_2 \ell_3}}\delta_{\ell_1 \ell'_3}\delta_{\ell_2 \ell'_2}\delta_{\ell_3 \ell'_1} \nonumber \\
&+\frac{C^{\rm XZ}_{\ell_1}C^{\rm YX}_{\ell_2}C^{\rm ZY}_{\ell_3}}{f_{\rm sky}N_{\ell_1 \ell_2 \ell_3}}\delta_{\ell_1 \ell'_3}\delta_{\ell_2 \ell'_1}\delta_{\ell_3 \ell'_2},
\end{align}
with $N_{\ell_1 \ell_2 \ell_3}$, being the number of modes for the ($\ell_1, \ell_2, \ell_3$) triangle.\\

For our purpose, we considered the tSZ-kSZ-kSZ bispectrum, thus we have two point that are identical. Considering that the expectation of tSZ and kSZ cross-correlation power spectrum, $C^{\rm tSZ,kSZ}_\ell$, is zero, we can safely neglect this term in the computation of uncertainties.
In this context Eq.~\ref{eqvar} reduces to
\begin{align}
\label{eqvar2}
<b_{\ell_1\ell_2\ell_3},b_{\ell'_1\ell'_2\ell'_3}> =& \frac{C^{\rm XX}_{\ell_1}C^{\rm XX}_{\ell_2}C^{\rm ZZ}_{\ell_3}}{f_{\rm sky}N_{\ell_1,\ell_2,\ell_3}}\delta_{\ell_1 \ell'_1}\delta_{\ell_2 \ell'_2}\delta_{\ell_3 \ell'_3} \nonumber \\
&+\frac{C^{\rm XX}_{\ell_1}C^{\rm XX}_{\ell_2}C^{\rm ZZ}_{\ell_3}}{f_{\rm sky}N_{\ell_1 \ell_2 \ell_3}}\delta_{\ell_1 \ell'_2}\delta_{\ell_2 \ell'_1}\delta_{\ell_3 \ell'_3} \nonumber \\
\end{align}
If we have $\ell_1 = \ell_2$ then 
\begin{align}
<b_{\ell_1\ell_2\ell_3},b_{\ell_1\ell_2\ell_3}> = 2\frac{C^{\rm XX}_{\ell_1}C^{\rm XX}_{\ell_2}C^{\rm ZZ}_{\ell_3}}{f_{\rm sky}N_{\ell_1\ell_2\ell_3}},
\end{align}
and if we have $\ell_1 \neq \ell_2$ then
\begin{align}
<b_{\ell_1\ell_2\ell_3},b_{\ell_1\ell_2\ell_3}> = \frac{C^{\rm XX}_{\ell_1}C^{\rm XX}_{\ell_2}C^{\rm ZZ}_{\ell_3}}{f_{\rm sky}N_{\ell_1\ell_2\ell_3}},
\end{align}

\subsection{Foreground contamination}

The two main foregrounds contamination in SZ analyses are the CIB and extra-galactic radio sources.\\
CIB leakage in component separated maps is composed by high-z CIB sources that present a spectral behavior than the thermal dust in our galaxy that drive the computation of weights for ILC based methods. As a consequence, CIB in such maps is almost gaussian and will not significantly bias the results.
The level of CIB in the final bispectrum can be estimated through the different shape of CIB and SZ bispectra \cite{lac14}.\\
Radio sources with a significant flux are in small number on the microwave sky. It in has been shown in tSZ bispectra analyses \citep{planckSZS} that the results is not biased by radio sources, that would produce a negative bias due to the way radio sources leak in tSZ maps \citep{hur13}.\\

\section{Forecasts}
\label{forecast}
\subsection{Cosmic variance limited experiment}

In a first step we estimated the expected S/N is we are just limited by the sky containing tSZ, kSZ, and CMB anisotropies.
Indeed tSZ can be extracted from other components \citep{hur13}, but the kSZ signal can not be distinguished from the CMB primary anisotropies or secondary anisotropies that follows the CMB black body emission law. We also consider $f_{\rm sky}=0.5$. This choice of a small sky fraction is motivated to avoid foreground residuals contamination that contaminates tSZ and CMB signals.\\
We present on Fig.~\ref{figsnr} the expected cumulative S/N as a function of the multipole $\ell$. We observed that the tSZ-kSZ-kSZ can be detected for scales above $\ell \simeq 3000$. 

\subsection{Realistic CMB experiments}

The Planck experiment have produced a cosmic-variance limited measurement of the CMB temperature angular power spectrum \citep{PlanckPS}. However, secondary anisotropies (such as the tSZ effect) measurement are still dominated by instrumental noise. In this section we present the expected signal to noise from the Planck experiment and from a future COrE+\footnote{\url{http://hdl.handle.net/11299/169642}} like CMB mission.

\subsubsection{Planck-like experiment}

\begin{figure}[!th]
\begin{center}
\includegraphics[scale=0.2,trim = 0cm 0cm 0cm 1.3cm, clip]{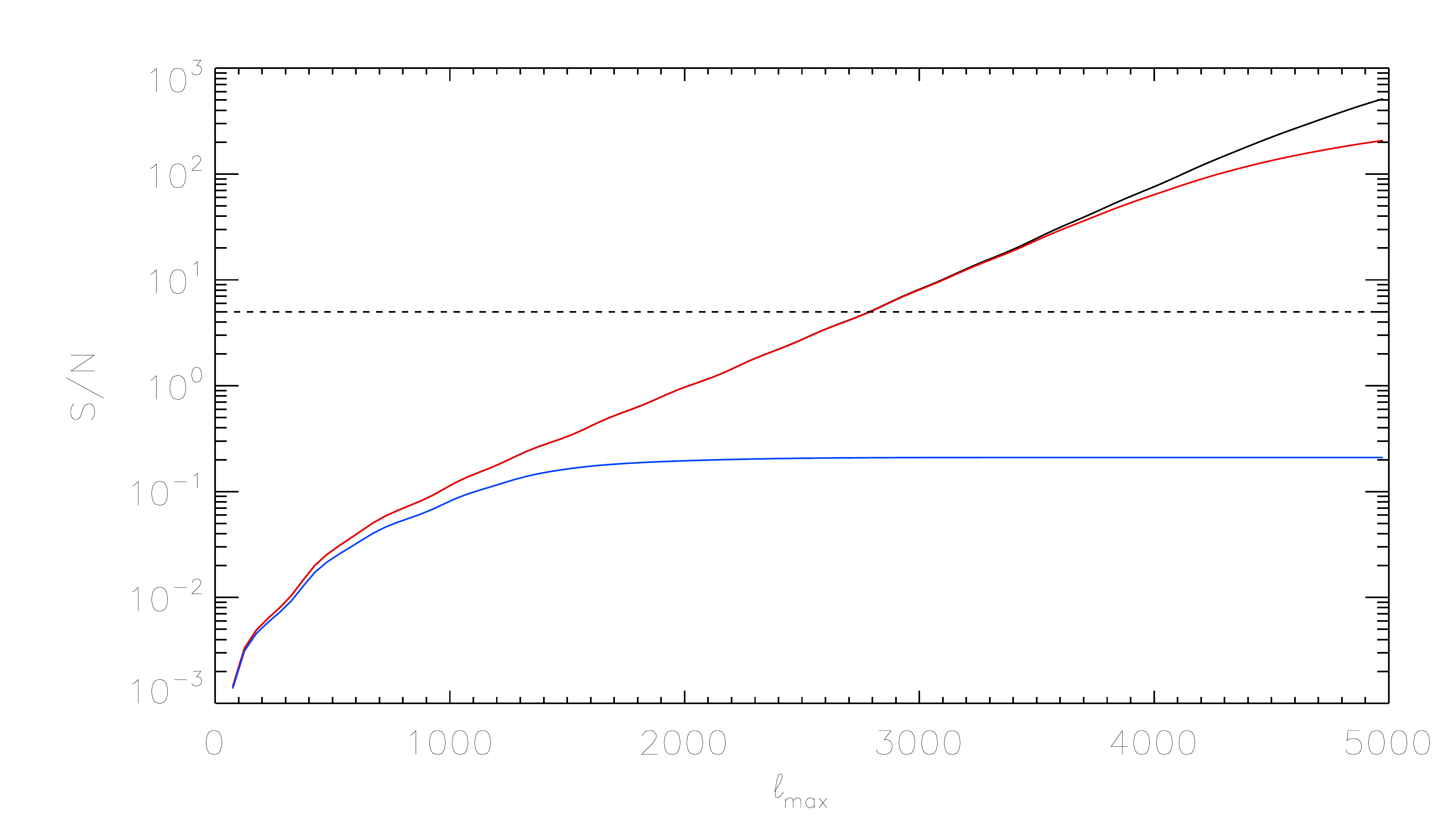}
\caption{Cumulative S/N for the tSZ-kSZ-kSZ bispectrum as a function of $\ell$ for a cosmic variance limited experiment (black line), the Planck experiment (blue line), and a COrE+ like experiment (red line). The dashed line shows the 5 $\sigma$ level.}.
\label{figsnr}
\end{center}
\end{figure}

To estimate the expected signal-to-noise ratio using data from a {\it Planck}-like mission, we used the noise level from {\it Planck} public CMB \citep{planckcmb} and tSZ \citep{planckszmap} maps assuming $f_{\rm sky}=0.5$.
For both CMB and tSZ maps we used the total measured power-spectra to estimate the noise level in the tSZ-kSZ-kSZ bispectrum. Consequently, this estimation of the noise level accounts for CIB and point sources contamination in both tSZ and CMB maps.\\
In Fig.~\ref{figsnr}, we present the expected signal-to-noise as a function of the maximum $\ell$ considered.
We observe that Planck is expected to achieve a 0.2 $\sigma$ measurement of the tSZ-kSZ-kSZ bispectrum.
This rules out the possibility to extract this signal from the Planck data.
The main limitations are the noise level in tSZ and CMB map, residual from other astrophysical components (mainly the CIB for the tSZ map), and the angular resolution of Planck maps, 5 arcmin FWHM for the CMB maps and 10 arcmin for the tSZ maps. The CMB also strongly limits the sensitivity at low-$\ell$. On Fig.~\ref{figsnr}, we can see low-$\ell$ oscillations that correspond to the CMB acoustic oscillation pics contributing to the noise for the tSZ-kSZ-kSZ bispectrum.

\subsubsection{Future CMB experiments}

\begin{figure*}[!th]
\begin{center}
\includegraphics[scale=0.4]{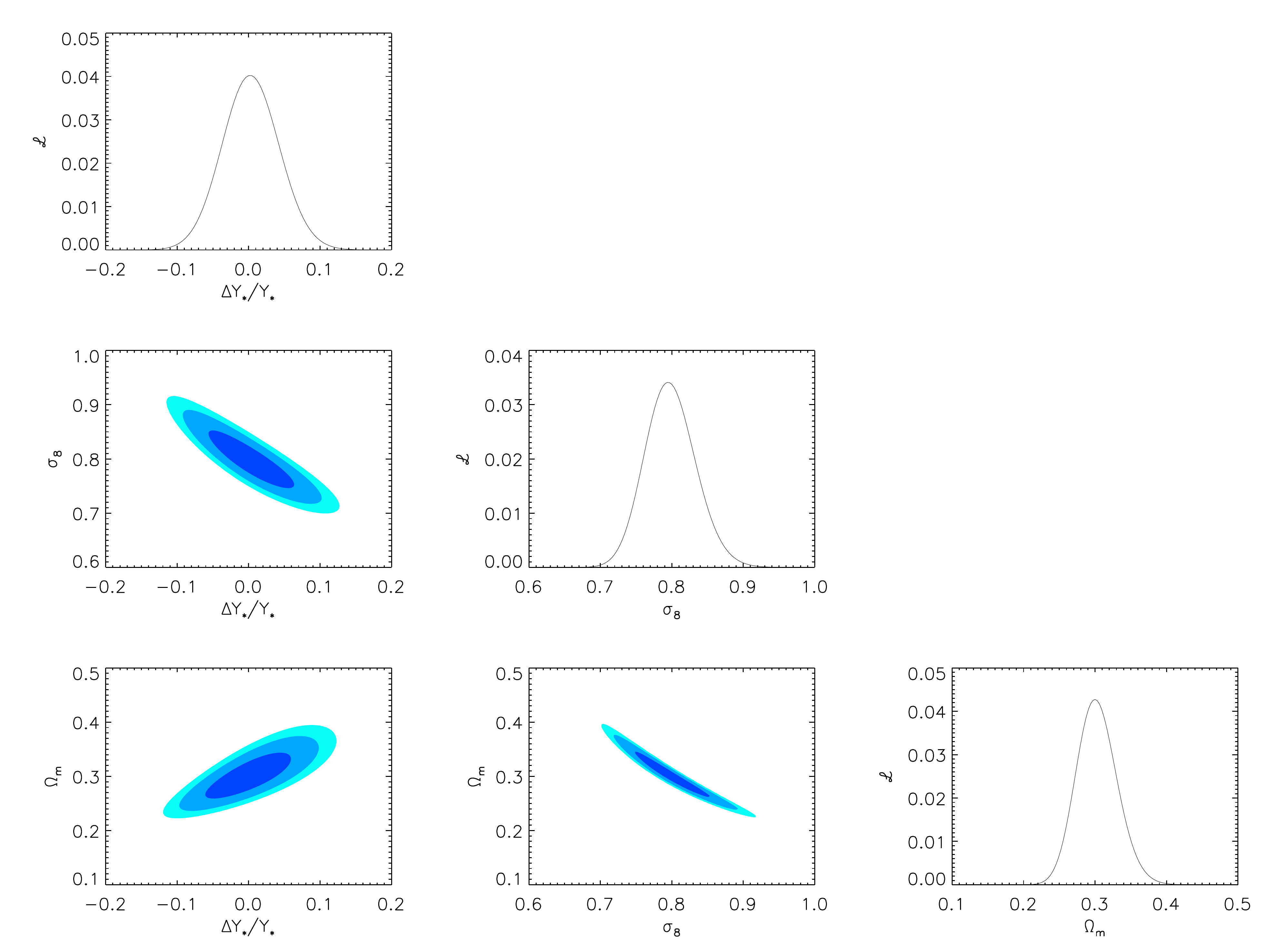}
\caption{Likelihood for combined tSZ power spectrum, bispectrum, and tSZ-kSZ-kSZ bispectrum as a function of $Y_\star$, $\sigma_8$, and $\Omega_{\rm m}$. Dark blue, blue, and light blue contours indicates the 1, 2, and 3 $\sigma$ levels.}.
\label{figlike}
\end{center}
\end{figure*}
We now consider a future CMB spacecraft mission assuming specificities (frequencies, noise level par detector, number of detectors, beams) for a COrE+ like experiment, and a sky coverage $f_{\rm sky}=0.5$.
For such future experiment, we don't have directly access to CMB or tSZ maps noise level. Thus, we computed the expected noise level in component separated maps obtained through linear combination of multi-frequency intensity maps. The optimal noise level, $V_i$, for a single astrophysical component is given by,
\begin{align}
V_i = \left({\mathbf F}_i {\cal C}^{-1}_N {\mathbf F}^T_i \right)^{-1},
\end{align}
where ${\mathbf F}_i$ is the component spectral behavior in the frequency channels of the experiment, and ${\cal C}^{-1}_N$ is the instrumental noise covariance matrix. We assume that ${\cal C}^{-1}_N$ is diagonal, and that multiple detector at a given frequency have uncorrelated noise.\\
However, there is several components on the sky. Thus, we use the following equation
\begin{align}
{\cal V} = \left({\cal F} {\cal C}^{-1}_N {\cal F}^T \right)^{-1},
\end{align}
where ${\cal F}$ is a rectangular matrix containing the astrophysical component spectral behavior. 
In this analysis, we considered:
\begin{itemize}
\item[$\bullet$] the tSZ effect, 
\item[$\bullet$] the CMB, 
\item[$\bullet$] one thermal dust component, that follows a modified black-body spectral energy distribution (SED) with a temperature, $T_d = 20 K$, and spectral index $\beta_d = 1.6$, 
\item[$\bullet$] one radio component following a $\nu^{\alpha_r}$ SED with a spectral index $\alpha_r = -1$, 
\item[$\bullet$] the CO component, 
\item[$\bullet$] the spinning dust component.
\end{itemize}
Additionally, some components, such as the cosmic-infrared-background (CIB), cannot be modeled with a single spectral law and contribute as a partially correlated component from frequency to frequency. We thus model the CIB contribution to the final variance as 
\begin{align}
{\cal V} = \left({\cal F} \left[ {\cal C}_N + {\cal C}_{\rm CIB} \right]^{-1} {\cal F}^T \right)^{-1},
\end{align}
where ${\cal C}_{\rm CIB}$ is the CIB covariance matrix. We computed the CIB covariance matrix using the model presented in \citep{planckszcib}. For a more realistic estimation we performed the noise estimation as a function of the multipole $\ell$ to have an estimate of the noise power-spectrum in CMB and tSZ maps.
We verified that this approach is realistic by applying it to the Planck mission specificities and comparing to the noise level observed in Planck tSZ and CMB public maps.

In Fig.~\ref{figsnr}, we present the expected tSZ-kSZ-kSZ bispectrum S/N for a COrE+ like experiment. We observe that the tSZ-kSZ-kSZ S/N is dominated by the cosmic variance up to $\ell = 4000$. We can expect a detection up to $200\, \sigma$ when neglecting potential systematic effect in the CMB and tSZ maps.
We also stress that our estimation of the tSZ-kSZ-kSZ bispectrum may under-estimate the real signal at high-$\ell$, where the internal dynamics of the gaz inside galaxy clusters will most likely add extra power on small scales.

In Fig.~\ref{figlike}, we present the expected constraints on $\Omega_{\rm m}$, $\sigma_8$, and $Y_\star$ when combining the measurement of tSZ power spectrum, bispectrum, and tSZ-kSZ-kSZ bispectrum.\\
The Likelihood function is computed as follows,
\begin{align}
{\cal L} \propto {\rm exp} \left( - \left[{\bf D} - {\bf M}(\sigma_8,\Omega_{\rm m}, Y_\star)\right]^{\rm T} {\cal C}_{\rm SZ} \left[ {\bf D} - {\bf M}(\sigma_8,\Omega_{\rm m}, Y_\star) \right]  \right),
\end{align}
Where ${\bf D}$ is a vector containing the tSZ-tSZ-tSZ and tSZ-kSZ-kSZ bispectra for our fiducial model ($\sigma_8 = 0.8$, $\Omega_m = 0.3$, $\Delta Y_\star/Y_\star = 0$), {\bf M} is a vector containing the two bispectra for parameters ($\sigma_8$, $\Omega_m$, $\Delta Y_\star/Y_\star$), and ${\cal C}_{\rm SZ}$ is the covariance matrix of the two bispectra in the weakly non-gaussian limit (see Eq.~\ref{eqvar}). The correlation between the tSZ-tSZ-tSZ and tSZ-kSZ-kSZ bispectra in the weakly non-gaussian limit is proportional to $\left(C^{\rm tSZ,kSZ}_\ell \right)^2$, thus contraints from the tSZ-tSZ-SZ bispectrum and from the tSZ-kSZ-kSZ bispectrum can be considered as independent. 
Figure~\ref{figlike} shows the expected constraints when $\sigma_8$, $\Omega_{\rm m}$, and $Y_\star$ are allowed to vary, we marginalized over $H_0 = 67.8 \pm 0.9$ km/s/Mpc and fixed all other parameters.

We observed that $\sigma_8$ and $\Omega_{\rm m}$ present a high degree of degeneracy, as our three probes present similar degeneracies for this two parameters. However, we observed that we can achieve a precision of 4\% on $Y_\star$, without any external prior. Such approach will allows to calibrate the tSZ scaling relation without the need of X-ray hydrostatic mass.

\section{Conclusion and discussion}

We have proposed a new method to detect kSZ effect using future high-resolution CMB experiments.
This method presents the advantage to be sensitive to the galaxy cluster velocity dispersion without bias from CMB auto-correlation or from kSZ effect produce by the diffuse baryonic gas at high redshift. This method also allows to constraints the velocity field without selection function. 
By comparison, the kSZ angular power spectrum measured by \citet{geo15}, is sensitive to the total kSZ power-spectrum. The constraints achieved on the kSZ power spectrum by \citet{cra14} when combining tSZ power spectrum and bispectrum is an indirect constraints, and is affected by a strong cosmic variance, as it measures the tSZ-tSZ-tSZ and tSZ-kSZ-kSZ bispectra as a single quantity.\\
The method proposed in the present paper, rely on a direct measurement of the tSZ-kSZ-kSZ bispectrum after a separation of the tSZ and kSZ signals. Consequently, it allows to obtain a model-independent and low-cosmic-variance estimation of the tSZ-kSZ-kSZ bispectrum.

We have presented a complete modeling of the tSZ-kSZ-kSZ bispectrum and deduced the associated cosmological parameter dependancies. We derived the dependencies of the tSZ-kSZ-kSZ and tSZ bispectra with respect to cosmological parameters. Previous works have also discuss the tSZ bispectrum scaling with cosmological parameters \citep[see e.g.,]{bat12,cra14}, and found slightly different scaling. The cosmological dependancies of the tSZ angular power spectrum is highly dependent on the weighting given to each halo. As a consequence, differences on the masse-observable relation will significantly affects the bispectrum scaling with respect to cosmological parameters. This also explains the scale dependance of the bispectrum scalings, indeed different angular scales receive contribution from halos at différent redshifts and masses.

We also demonstrated that future experiments, will be sensitive to the tSZ-kSZ-kSZ cross-correlation bispectrum up to $200\, \sigma$.\\
We demonstrated that the tSZ-kSZ-kSZ bispectrum can be combined with the tSZ power spectrum and bispectrum to set tight constraints (4\%) on the $Y-M$ relation calibration and thus on the hydrostatic mass bias. This will enable the possibility to set cosmological constraints without the need of prior on the hydrostatic mass bias. Which is a crucial step considering that the hydrostatic mass-bias is the main limitation for cosmological parameter estimation from tSZ surveys.

\section*{Acknowledgements}
\thanks{The author thanks F.Lacasa for useful discussions. We acknowledge the support of the French \emph{Agence Nationale de la Recherche} under grant ANR-11-BD56-015. 

\bibliographystyle{aa}
\bibliography{sz_bispectre.bib}

\begin{thebibliography}{55}
\expandafter\ifx\csname natexlab\endcsname\relax\def\natexlab#1{#1}\fi

\bibitem[{{Adam} {et~al.}(2016){Adam}, {Bartalucci}, {Pratt}, {Ade},
  {Andr{\'e}}, {Arnaud}, {Beelen}, {Beno{\^i}t}, {Bideaud}, {Billot},
  {Bourdin}, {Bourrion}, {Calvo}, {Catalano}, {Coiffard}, {Comis}, {D'Addabbo},
  {De Petris}, {D{\'e}sert}, {D{\'e}mocl{\`e}s}, {Doyle}, {Egami}, {Ferrari},
  {Goupy}, {Kramer}, {Lagache}, {Leclercq}, {Mac{\'{\i}}as-P{\'e}rez},
  {Maurogordato}, {Mauskopf}, {Mayet}, {Monfardini}, {Mroczkowski}, {Pajot},
  {Pascale}, {Perotto}, {Pisano}, {Pointecouteau}, {Ponthieu}, {Rev{\'e}ret},
  {Ritacco}, {Rodriguez}, {Romero}, {Ruppin}, {Schuster}, {Sievers},
  {Triqueneaux}, {Tucker}, {Zemcov}, \& {Zylka}}]{ada16}
{Adam}, R., {Bartalucci}, I., {Pratt}, G.~W., {et~al.} 2016, ArXiv e-prints

\bibitem[{{Addison} {et~al.}(2013){Addison}, {Dunkley}, \& {Bond}}]{add13}
{Addison}, G.~E., {Dunkley}, J., \& {Bond}, J.~R. 2013, \mnras, 436, 1896

\bibitem[{{Arnaud} {et~al.}(2010){Arnaud}, {Pratt}, {Piffaretti},
  {B{\"o}hringer}, {Croston}, \& {Pointecouteau}}]{arn10}
{Arnaud}, M., {Pratt}, G.~W., {Piffaretti}, R., {et~al.} 2010, \aap, 517, A92

\bibitem[{{Bhattacharya} {et~al.}(2012{\natexlab{a}}){Bhattacharya}, {Nagai},
  {Shaw}, {Crawford}, \& {Holder}}]{bha12}
{Bhattacharya}, S., {Nagai}, D., {Shaw}, L., {Crawford}, T., \& {Holder}, G.~P.
  2012{\natexlab{a}}, \apj, 760, 5

\bibitem[{{Bhattacharya} {et~al.}(2012{\natexlab{b}}){Bhattacharya}, {Nagai},
  {Shaw}, {Crawford}, \& {Holder}}]{bat12}
{Bhattacharya}, S., {Nagai}, D., {Shaw}, L., {Crawford}, T., \& {Holder}, G.~P.
  2012{\natexlab{b}}, \apj, 760, 5

\bibitem[{{Birkinshaw}(1999)}]{bir99}
{Birkinshaw}, M. 1999, \physrep, 310, 97

\bibitem[{{Bleem} {et~al.}(2015){Bleem}, {Stalder}, {de Haan}, {Aird}, {Allen},
  {Applegate}, {Ashby}, {Bautz}, {Bayliss}, {Benson}, {Bocquet}, {Brodwin},
  {Carlstrom}, {Chang}, {Chiu}, {Cho}, {Clocchiatti}, {Crawford}, {Crites},
  {Desai}, {Dietrich}, {Dobbs}, {Foley}, {Forman}, {George}, {Gladders},
  {Gonzalez}, {Halverson}, {Hennig}, {Hoekstra}, {Holder}, {Holzapfel},
  {Hrubes}, {Jones}, {Keisler}, {Knox}, {Lee}, {Leitch}, {Liu}, {Lueker},
  {Luong-Van}, {Mantz}, {Marrone}, {McDonald}, {McMahon}, {Meyer}, {Mocanu},
  {Mohr}, {Murray}, {Padin}, {Pryke}, {Reichardt}, {Rest}, {Ruel}, {Ruhl},
  {Saliwanchik}, {Saro}, {Sayre}, {Schaffer}, {Schrabback}, {Shirokoff},
  {Song}, {Spieler}, {Stanford}, {Staniszewski}, {Stark}, {Story}, {Stubbs},
  {Vanderlinde}, {Vieira}, {Vikhlinin}, {Williamson}, {Zahn}, \&
  {Zenteno}}]{ble15}
{Bleem}, L.~E., {Stalder}, B., {de Haan}, T., {et~al.} 2015, \apjs, 216, 27

\bibitem[{{Carlstrom} {et~al.}(2002){Carlstrom}, {Holder}, \& {Reese}}]{car02}
{Carlstrom}, J.~E., {Holder}, G.~P., \& {Reese}, E.~D. 2002, \araa, 40, 643

\bibitem[{{Cole} \& {Kaiser}(1988)}]{col88}
{Cole}, S. \& {Kaiser}, N. 1988, \mnras, 233, 637

\bibitem[{{Crawford} {et~al.}(2014){Crawford}, {Schaffer}, {Bhattacharya},
  {Aird}, {Benson}, {Bleem}, {Carlstrom}, {Chang}, {Cho}, {Crites}, {de Haan},
  {Dobbs}, {Dudley}, {George}, {Halverson}, {Holder}, {Holzapfel}, {Hoover},
  {Hou}, {Hrubes}, {Keisler}, {Knox}, {Lee}, {Leitch}, {Lueker}, {Luong-Van},
  {McMahon}, {Mehl}, {Meyer}, {Millea}, {Mocanu}, {Mohr}, {Montroy}, {Padin},
  {Plagge}, {Pryke}, {Reichardt}, {Ruhl}, {Sayre}, {Shaw}, {Shirokoff},
  {Spieler}, {Staniszewski}, {Stark}, {Story}, {van Engelen}, {Vanderlinde},
  {Vieira}, {Williamson}, \& {Zahn}}]{cra14}
{Crawford}, T.~M., {Schaffer}, K.~K., {Bhattacharya}, S., {et~al.} 2014, \apj,
  784, 143

\bibitem[{{de Haan} {et~al.}(2016){de Haan}, {Benson}, {Bleem}, {Allen},
  {Applegate}, {Ashby}, {Bautz}, {Bayliss}, {Bocquet}, {Brodwin}, {Carlstrom},
  {Chang}, {Chiu}, {Cho}, {Clocchiatti}, {Crawford}, {Crites}, {Desai},
  {Dietrich}, {Dobbs}, {Doucouliagos}, {Foley}, {Forman}, {Garmire}, {George},
  {Gladders}, {Gonzalez}, {Gupta}, {Halverson}, {Hlavacek-Larrondo},
  {Hoekstra}, {Holder}, {Holzapfel}, {Hou}, {Hrubes}, {Huang}, {Jones},
  {Keisler}, {Knox}, {Lee}, {Leitch}, {von der Linden}, {Luong-Van}, {Mantz},
  {Marrone}, {McDonald}, {McMahon}, {Meyer}, {Mocanu}, {Mohr}, {Murray},
  {Padin}, {Pryke}, {Rapetti}, {Reichardt}, {Rest}, {Ruel}, {Ruhl},
  {Saliwanchik}, {Saro}, {Sayre}, {Schaffer}, {Schrabback}, {Shirokoff},
  {Song}, {Spieler}, {Stalder}, {Stanford}, {Staniszewski}, {Stark}, {Story},
  {Stubbs}, {Vanderlinde}, {Vieira}, {Vikhlinin}, {Williamson}, \&
  {Zenteno}}]{haa16}
{de Haan}, T., {Benson}, B.~A., {Bleem}, L.~E., {et~al.} 2016, ArXiv e-prints

\bibitem[{{DeDeo} {et~al.}(2005){DeDeo}, {Spergel}, \& {Trac}}]{ded05}
{DeDeo}, S., {Spergel}, D.~N., \& {Trac}, H. 2005, ArXiv Astrophysics e-prints

\bibitem[{{Diaferio} {et~al.}(2000){Diaferio}, {Sunyaev}, \& {Nusser}}]{dia00}
{Diaferio}, A., {Sunyaev}, R.~A., \& {Nusser}, A. 2000, \apjl, 533, L71

\bibitem[{{Diego} \& {Majumdar}(2004)}]{die04}
{Diego}, J.~M. \& {Majumdar}, S. 2004, \mnras, 352, 993

\bibitem[{{Dor{\'e}} {et~al.}(2004){Dor{\'e}}, {Hennawi}, \& {Spergel}}]{dor04}
{Dor{\'e}}, O., {Hennawi}, J.~F., \& {Spergel}, D.~N. 2004, \apj, 606, 46

\bibitem[{{George} {et~al.}(2015){George}, {Reichardt}, {Aird}, {Benson},
  {Bleem}, {Carlstrom}, {Chang}, {Cho}, {Crawford}, {Crites}, {de Haan},
  {Dobbs}, {Dudley}, {Halverson}, {Harrington}, {Holder}, {Holzapfel}, {Hou},
  {Hrubes}, {Keisler}, {Knox}, {Lee}, {Leitch}, {Lueker}, {Luong-Van},
  {McMahon}, {Mehl}, {Meyer}, {Millea}, {Mocanu}, {Mohr}, {Montroy}, {Padin},
  {Plagge}, {Pryke}, {Ruhl}, {Schaffer}, {Shaw}, {Shirokoff}, {Spieler},
  {Staniszewski}, {Stark}, {Story}, {van Engelen}, {Vanderlinde}, {Vieira},
  {Williamson}, \& {Zahn}}]{geo15}
{George}, E.~M., {Reichardt}, C.~L., {Aird}, K.~A., {et~al.} 2015, \apj, 799,
  177

\bibitem[{{Hand} {et~al.}(2012){Hand}, {Addison}, {Aubourg}, {Battaglia},
  {Battistelli}, {Bizyaev}, {Bond}, {Brewington}, {Brinkmann}, {Brown}, {Das},
  {Dawson}, {Devlin}, {Dunkley}, {Dunner}, {Eisenstein}, {Fowler}, {Gralla},
  {Hajian}, {Halpern}, {Hilton}, {Hincks}, {Hlozek}, {Hughes}, {Infante},
  {Irwin}, {Kosowsky}, {Lin}, {Malanushenko}, {Malanushenko}, {Marriage},
  {Marsden}, {Menanteau}, {Moodley}, {Niemack}, {Nolta}, {Oravetz}, {Page},
  {Palanque-Delabrouille}, {Pan}, {Reese}, {Schlegel}, {Schneider}, {Sehgal},
  {Shelden}, {Sievers}, {Sif{\'o}n}, {Simmons}, {Snedden}, {Spergel}, {Staggs},
  {Swetz}, {Switzer}, {Trac}, {Weaver}, {Wollack}, {Yeche}, \&
  {Zunckel}}]{han12}
{Hand}, N., {Addison}, G.~E., {Aubourg}, E., {et~al.} 2012, Physical Review
  Letters, 109, 041101

\bibitem[{{Hasselfield} {et~al.}(2013){Hasselfield}, {Hilton}, {Marriage},
  {Addison}, {Barrientos}, {Battaglia}, {Battistelli}, {Bond}, {Crichton},
  {Das}, {Devlin}, {Dicker}, {Dunkley}, {D{\"u}nner}, {Fowler}, {Gralla},
  {Hajian}, {Halpern}, {Hincks}, {Hlozek}, {Hughes}, {Infante}, {Irwin},
  {Kosowsky}, {Marsden}, {Menanteau}, {Moodley}, {Niemack}, {Nolta}, {Page},
  {Partridge}, {Reese}, {Schmitt}, {Sehgal}, {Sherwin}, {Sievers}, {Sif{\'o}n},
  {Spergel}, {Staggs}, {Swetz}, {Switzer}, {Thornton}, {Trac}, \&
  {Wollack}}]{has13}
{Hasselfield}, M., {Hilton}, M., {Marriage}, T.~A., {et~al.} 2013, \jcap, 7,
  008

\bibitem[{{Hern{\'a}ndez-Monteagudo} {et~al.}(2015){Hern{\'a}ndez-Monteagudo},
  {Ma}, {Kitaura}, {Wang}, {G{\'e}nova-Santos}, {Mac{\'{\i}}as-P{\'e}rez}, \&
  {Herranz}}]{her15}
{Hern{\'a}ndez-Monteagudo}, C., {Ma}, Y.-Z., {Kitaura}, F.~S., {et~al.} 2015,
  Physical Review Letters, 115, 191301

\bibitem[{{Hill} \& {Sherwin}(2013)}]{hil13}
{Hill}, J.~C. \& {Sherwin}, B.~D. 2013, \prd, 87, 023527

\bibitem[{{Ho} {et~al.}(2009){Ho}, {Dedeo}, \& {Spergel}}]{ho09}
{Ho}, S., {Dedeo}, S., \& {Spergel}, D. 2009, ArXiv e-prints

\bibitem[{{Hurier} {et~al.}(2013){Hurier}, {Hildebrandt}, \&
  {Macias-Perez}}]{hur13}
{Hurier}, G., {Hildebrandt}, S.~R., \& {Macias-Perez}, J.~F. 2013, ArXiv
  e-prints

\bibitem[{{Kitaura} {et~al.}(2012){Kitaura}, {Angulo}, {Hoffman}, \&
  {Gottl{\"o}ber}}]{kit12}
{Kitaura}, F.-S., {Angulo}, R.~E., {Hoffman}, Y., \& {Gottl{\"o}ber}, S. 2012,
  \mnras, 425, 2422

\bibitem[{{Komatsu} \& {Kitayama}(1999)}]{kom99}
{Komatsu}, E. \& {Kitayama}, T. 1999, \apjl, 526, L1

\bibitem[{{Komatsu} \& {Seljak}(2001)}]{kom01}
{Komatsu}, E. \& {Seljak}, U. 2001, \mnras, 327, 1353

\bibitem[{{Komatsu} \& {Seljak}(2002)}]{kom02}
{Komatsu}, E. \& {Seljak}, U. 2002, \mnras, 336, 1256

\bibitem[{{Lacasa}(2014)}]{lac14}
{Lacasa}, F. 2014, ArXiv e-prints

\bibitem[{{Mantz} {et~al.}(2015){Mantz}, {von der Linden}, {Allen},
  {Applegate}, {Kelly}, {Morris}, {Rapetti}, {Schmidt}, {Adhikari}, {Allen},
  {Burchat}, {Burke}, {Cataneo}, {Donovan}, {Ebeling}, {Shandera}, \&
  {Wright}}]{man15}
{Mantz}, A.~B., {von der Linden}, A., {Allen}, S.~W., {et~al.} 2015, \mnras,
  446, 2205

\bibitem[{{Mather} {et~al.}(1999){Mather}, {Fixsen}, {Shafer}, {Mosier}, \&
  {Wilkinson}}]{mat99}
{Mather}, J.~C., {Fixsen}, D.~J., {Shafer}, R.~A., {Mosier}, C., \&
  {Wilkinson}, D.~T. 1999, \apj, 512, 511

\bibitem[{{Mo} \& {White}(1996)}]{mo96}
{Mo}, H.~J. \& {White}, S.~D.~M. 1996, \mnras, 282, 347

\bibitem[{{Nagai} {et~al.}(2007){Nagai}, {Kravtsov}, \& {Vikhlinin}}]{nag07}
{Nagai}, D., {Kravtsov}, A.~V., \& {Vikhlinin}, A. 2007, \apj, 668, 1

\bibitem[{{Navarro} {et~al.}(1997){Navarro}, {Frenk}, \& {White}}]{nav97}
{Navarro}, J.~F., {Frenk}, C.~S., \& {White}, S.~D.~M. 1997, \apj, 490, 493

\bibitem[{{Ostriker} \& {Vishniac}(1986)}]{ost86}
{Ostriker}, J.~P. \& {Vishniac}, E.~T. 1986, \apjl, 306, L51

\bibitem[{{Peebles}(1980)}]{pee80}
{Peebles}, P.~J.~E. 1980, {The large-scale structure of the universe}

\bibitem[{{Planck Collaboration 2015 Results I}(2015)}]{planckover}
{Planck Collaboration 2015 Results I}. 2015, ArXiv e-prints

\bibitem[{{Planck Collaboration 2015 Results IX}(2015)}]{planckcmb}
{Planck Collaboration 2015 Results IX}. 2015, ArXiv e-prints

\bibitem[{{Planck Collaboration 2015 Results XIII}(2015)}]{planckcosmo}
{Planck Collaboration 2015 Results XIII}. 2015, ArXiv e-prints

\bibitem[{{Planck Collaboration 2015 Results XXII}(2015)}]{planckszmap}
{Planck Collaboration 2015 Results XXII}. 2015, ArXiv e-prints

\bibitem[{{Planck Collaboration 2015 Results XXIII}(2015)}]{planckszcib}
{Planck Collaboration 2015 Results XXIII}. 2015, ArXiv e-prints

\bibitem[{{Planck Collaboration 2015 Results XXIV}(2015)}]{planckszcount}
{Planck Collaboration 2015 Results XXIV}. 2015, ArXiv e-prints

\bibitem[{{Planck Collaboration 2015 Results XXVII}(2015)}]{planckpsz2}
{Planck Collaboration 2015 Results XXVII}. 2015, ArXiv e-prints

\bibitem[{{Planck Collaboration early results. XI}(2011)}]{planckSL}
{Planck Collaboration early results. XI}. 2011, \aap, 536, A11

\bibitem[{{Planck collaboration results XV}(2013)}]{PlanckPS}
{Planck collaboration results XV}. 2013, ArXiv e-prints

\bibitem[{{Planck Collaboration results XXI}(2013)}]{planckSZS}
{Planck Collaboration results XXI}. 2013, ArXiv e-prints

\bibitem[{{Planck Collaboration results XXIX}(2013)}]{PlanckSZC}
{Planck Collaboration results XXIX}. 2013, ArXiv e-prints

\bibitem[{{Remazeilles} {et~al.}(2011){Remazeilles}, {Delabrouille}, \&
  {Cardoso}}]{rem11}
{Remazeilles}, M., {Delabrouille}, J., \& {Cardoso}, J.-F. 2011, \mnras, 410,
  2481

\bibitem[{{Sayers} {et~al.}(2013){Sayers}, {Mroczkowski}, {Zemcov}, {Korngut},
  {Bock}, {Bulbul}, {Czakon}, {Egami}, {Golwala}, {Koch}, {Lin}, {Mantz},
  {Molnar}, {Moustakas}, {Pierpaoli}, {Rawle}, {Reese}, {Rex}, {Shitanishi},
  {Siegel}, \& {Umetsu}}]{say13}
{Sayers}, J., {Mroczkowski}, T., {Zemcov}, M., {et~al.} 2013, \apj, 778, 52

\bibitem[{{Shaw} {et~al.}(2012){Shaw}, {Rudd}, \& {Nagai}}]{sha12}
{Shaw}, L.~D., {Rudd}, D.~H., \& {Nagai}, D. 2012, \apj, 756, 15

\bibitem[{{Sievers} {et~al.}(2013){Sievers}, {Hlozek}, {Nolta}, {Acquaviva},
  {Addison}, {Ade}, {Aguirre}, {Amiri}, {Appel}, {Barrientos}, {Battistelli},
  {Battaglia}, {Bond}, {Brown}, {Burger}, {Calabrese}, {Chervenak}, {Crichton},
  {Das}, {Devlin}, {Dicker}, {Bertrand Doriese}, {Dunkley}, {D{\"u}nner},
  {Essinger-Hileman}, {Faber}, {Fisher}, {Fowler}, {Gallardo}, {Gordon},
  {Gralla}, {Hajian}, {Halpern}, {Hasselfield}, {Hern{\'a}ndez-Monteagudo},
  {Hill}, {Hilton}, {Hilton}, {Hincks}, {Holtz}, {Huffenberger}, {Hughes},
  {Hughes}, {Infante}, {Irwin}, {Jacobson}, {Johnstone}, {Baptiste Juin},
  {Kaul}, {Klein}, {Kosowsky}, {Lau}, {Limon}, {Lin}, {Louis}, {Lupton},
  {Marriage}, {Marsden}, {Martocci}, {Mauskopf}, {McLaren}, {Menanteau},
  {Moodley}, {Moseley}, {Netterfield}, {Niemack}, {Page}, {Page}, {Parker},
  {Partridge}, {Plimpton}, {Quintana}, {Reese}, {Reid}, {Rojas}, {Sehgal},
  {Sherwin}, {Schmitt}, {Spergel}, {Staggs}, {Stryzak}, {Swetz}, {Switzer},
  {Thornton}, {Trac}, {Tucker}, {Uehara}, {Visnjic}, {Warne}, {Wilson},
  {Wollack}, {Zhao}, \& {Zunckel}}]{sie13}
{Sievers}, J.~L., {Hlozek}, R.~A., {Nolta}, M.~R., {et~al.} 2013, \jcap, 10,
  060

\bibitem[{{Soergel} {et~al.}(2016){Soergel}, {Flender}, {Story}, {Bleem},
  {Giannantonio}, {Efstathiou}, {Rykoff}, {Benson}, {Crawford}, {Dodelson},
  {Habib}, {Heitmann}, {Holder}, {Jain}, {Rozo}, {Saro}, {Weller}, {Abdalla},
  {Allam}, {Annis}, {Armstrong}, {Benoit-L{\'e}vy}, {Bernstein}, {Carlstrom},
  {Carnero Rosell}, {Carrasco Kind}, {Castander}, {Chiu}, {Chown}, {Crocce},
  {Cunha}, {D'Andrea}, {da Costa}, {de Haan}, {Desai}, {Diehl}, {Dietrich},
  {Doel}, {Estrada}, {Evrard}, {Flaugher}, {Fosalba}, {Frieman}, {Gaztanaga},
  {Gruen}, {Gruendl}, {Holzapfel}, {Honscheid}, {James}, {Keisler}, {Kuehn},
  {Kuropatkin}, {Lahav}, {Lima}, {Marshall}, {McDonald}, {Melchior}, {Miller},
  {Miquel}, {Nord}, {Ogando}, {Omori}, {Plazas}, {Rapetti}, {Reichardt},
  {Romer}, {Roodman}, {Saliwanchik}, {Sanchez}, {Schubnell}, {Sevilla-Noarbe},
  {Sheldon}, {Smith}, {Soares-Santos}, {Sobreira}, {Stark}, {Suchyta},
  {Swanson}, {Tarle}, {Thomas}, {Vieira}, {Walker}, \& {Whitehorn}}]{soe16}
{Soergel}, B., {Flender}, S., {Story}, K.~T., {et~al.} 2016, \mnras, 461, 3172

\bibitem[{{Sugiyama} {et~al.}(2016){Sugiyama}, {Okumura}, \& {Spergel}}]{sug16}
{Sugiyama}, N.~S., {Okumura}, T., \& {Spergel}, D.~N. 2016, ArXiv e-prints

\bibitem[{{Sunyaev} \& {Zeldovich}(1972)}]{sun72}
{Sunyaev}, R.~A. \& {Zeldovich}, Y.~B. 1972, Comments on Astrophysics and Space
  Physics, 4, 173

\bibitem[{{Taburet} {et~al.}(2011){Taburet}, {Hern{\'a}ndez-Monteagudo},
  {Aghanim}, {Douspis}, \& {Sunyaev}}]{tab11}
{Taburet}, N., {Hern{\'a}ndez-Monteagudo}, C., {Aghanim}, N., {Douspis}, M., \&
  {Sunyaev}, R.~A. 2011, \mnras, 418, 2207

\bibitem[{{Tinker} {et~al.}(2008){Tinker}, {Kravtsov}, {Klypin}, {Abazajian},
  {Warren}, {Yepes}, {Gottl{\"o}ber}, \& {Holz}}]{tin08}
{Tinker}, J., {Kravtsov}, A.~V., {Klypin}, A., {et~al.} 2008, \apj, 688, 709

\bibitem[{{Wilson} {et~al.}(2012){Wilson}, {Sherwin}, {Hill}, {Addison},
  {Battaglia}, {Bond}, {Das}, {Devlin}, {Dunkley}, {D{\"u}nner}, {Fowler},
  {Gralla}, {Hajian}, {Halpern}, {Hilton}, {Hincks}, {Hlozek}, {Huffenberger},
  {Hughes}, {Kosowsky}, {Louis}, {Marriage}, {Marsden}, {Menanteau}, {Moodley},
  {Niemack}, {Nolta}, {Page}, {Partridge}, {Reese}, {Sehgal}, {Sievers},
  {Spergel}, {Staggs}, {Swetz}, {Switzer}, {Trac}, \& {Wollack}}]{wil12}
{Wilson}, M.~J., {Sherwin}, B.~D., {Hill}, J.~C., {et~al.} 2012, \prd, 86,
  122005

\end{thebibliography}

\end{document}